\documentstyle[aps,epsf]{revtex}

\newcommand{\eins}{\mbox{$1 \hspace{-1.0mm}  {\bf l}$}}
\newcommand{\be}{\begin{equation}}
\newcommand{\ee}{\end{equation}}
\newcommand{\bea}{\begin{eqnarray}}
\newcommand{\eea}{\end{eqnarray}}
\newcommand{\half}{\mbox{$\textstyle \frac{1}{2}$}}

\newcommand{\shalf}{\mbox{$\textstyle \frac{1}{\sqrt{2}}$}}
\newcommand{\ket}[1]{ | \, #1  \rangle}
\newcommand{\bra}[1]{ \langle #1 \,  |}

\newcommand{\abs}[1]{ | \, #1 \,  |}

\begin{document}
\draft\onecolumn
\title{Optimal eavesdropping in quantum cryptography with six states
}
\author{Dagmar~Bru\ss  }
\address{
ISI, Villa Gualino, Viale Settimio Severo 65, 10133 Torino, Italy }
\date{Received \today}
\maketitle
\begin{abstract}
A generalization of the quantum cryptographic protocol by Bennett and
Brassard is discussed,
 using three conjugate bases, i.e. six states. 
 By calculating the optimal mutual information between sender and eavesdropper 
 it is shown that this  scheme is safer against
eavesdropping on single qubits than the one based on two conjugate bases.
We also address the question  for
 a connection between
 the maximal classical correlation in a generalized Bell inequality
 and the
intersection of mutual informations between sender/receiver and
sender/eavesdropper.
\end{abstract}

In 1984 Bennett and Brassard~\cite{bb84} suggested a 
quantum cryptographic protocol,
in the following called BB84,
which enables two parties to establish a secret key, using principles
 of quantum mechanics.   In this scheme the sender of the quantum information,
 usually called Alice, transmits quantum bits in the basis 
 $\ket{0},\ket{1}$ or the conjugate
 basis $\ket{\bar 0},\ket{\bar 1}$,
  defined by 
  \bea
 \ket{\bar 0} & = &  \shalf (\ket{0}+\ket{1}) \nonumber \\
 \ket{\bar 1} & = &  \shalf (\ket{0}-\ket{1})
 \eea
   to the receiver Bob, who performs measurements in 
 these bases. After classical communication via a public channel 
 a secret key can be
 established by using only those cases in which the bases of Alice and Bob
 coincide.
 \par In this paper we want to discuss a 
 generalized scheme which is based on the
 use of three rather than two bases. The third one used in addition to
 the previous ones is denoted by 
 $\ket{\bar{\bar 0}},\ket{\bar{\bar 1}}$ and 
  defined by 
  \bea
 \ket{\bar{\bar 0}} & = &  \shalf (\ket{0}+i\ket{1}) \nonumber \\
 \ket{\bar{\bar 1}} & = &  \shalf (\ket{0}-i\ket{1}) \ .
 \eea
   In 
 the Bloch vector picture  a density matrix $\varrho$ is written as 
 $\varrho= \half (\eins +\vec s\cdot \vec \sigma)$,
  with $\vec s$ being the Bloch vector and $\vec \sigma$ the Pauli matrices. 
 The six states can be viewed as Bloch vectors pointing 
 along the positive and negative x-, y- and
 z-directions. 
 Alice sends one out of these, denoted as  $\ket{\psi^{in}}$,
 with equal probability.
 \par Such a scenario is a straightforward extension of the traditional protocol
 and its possibility
 has  been mentioned at various occasions \cite{private}. 
  Our main purpose is to point out that this generalized
  scheme is principally more
 secure than the one in~\cite{bb84}. This is due to the fact that the optimal 
 strategy
 an eavesdropper, traditionally called Eve, can design to gather information by
 performing some unitary transformation on the quantum bit in transit, gives
 her in our scenario less 
 information for a fixed disturbance of Bob's qubit. As Alice increases the set
 of inputs it is more difficult for Eve to learn something in transit.
 \par It was conjectured in 
 \cite{gihu} and shown in \cite{fupe}  that in the BB84 scenario
 the disturbance corresponding
 to Bob and Eve possessing the same information with respect to Alice exhibits
 a connection to the CHSH inequality. 
 After deriving Eve's optimal strategy we will ask whether  in the generalized
 protocol the crossing point between the two relevant mutual informations
 has a connection to a generalized Bell inequality 
 where Alice and Bob use the observables
 $\vec{a}_i\cdot \vec{\sigma}^a$ and 
 $\vec{b}_i\cdot \vec{\sigma}^b, \{i=1,2,...,n\}$, respectively,
 with the
 Bloch vectors $\vec{a}_i, \vec{b}_i$
 spanning not only a plane, but the  Bloch sphere.
 \par After this introduction and outline of the paper let us derive the
  eavesdropping strategy that is optimal 
 with respect to the mutual information between Alice and Eve, $I^{AE}$.
  We do not consider collective or coherent
 attacks, but only interaction with single qubits. The most general
 unitary transformation Eve can design is of the form  
 \bea
U\, \ket{0}\ket{X} & = & \sqrt{F}\ket{0}\ket{A}+\sqrt{1-F}\ket{1}
  \ket{B}
 \label{eq:00}\\
U\, \ket{1}\ket{X} & = &  
     \sqrt{F}\ket{1}\ket{ C}
     +  \sqrt{1-F}\ket{0}\ket{ D} \ .
  \label{eq:10}
\eea
The first qubit is the one sent to Bob and acted on by Eve.
Eve's initial state is $\ket{X}$,  
and $\ket{A},\ket{B},\ket{C},\ket{D}$  refer to her normalized states
after the interaction. 
It was shown in \cite{fupe1} that it
is sufficient for Eve to use two qubits in order to extract the maximal
information.
The fidelity of Bob's bit is $F$ and is taken to be in the interval
$1/2\leq F\leq 1$.
\par We assume Eve to be clever enough to treat all six possible states in the
same way (i.e. with same disturbance for Bob) - otherwise Alice
 and Bob could find
out about her existence by comparing error rates
in different bases. This assumption results in three
constraints which the scalar products of Eve's states have to fulfill:
\bea
\bra{B} D\rangle & = & 0  \ ,\nonumber \\
Re \bra{C} A\rangle & = & 2-\frac{1}{F} \ ,\nonumber \\
\bra{A} B\rangle +\bra{D} C\rangle  & = & 0 \ .
\label{constraint1}
\eea
Unitarity of the matrix U means
\be
\bra{A} D\rangle +\bra{B} C\rangle   =  0  \ .
\label{constraint2}
\ee
\par
The mutual information between Alice and Bob 
is given by
\be
I^{AB}=1+ D \log D+ (1-D) \log (1-D) \ ,
\ee 
where D is the disturbance of Bob's qubit, defined by
\be
D=1-F=1-\bra{\psi^{in}}\varrho^B\ket{\psi^{in}}  \ ,
\ee 
and $ \varrho^B$ is the right hand side of equations 
(\ref{eq:00}) and (\ref{eq:10}), traced over Eve's bits.
All logarithms are taken to base 2.
By construction Bob's disturbance is the same no matter which state was sent by
Alice.
The procedure to calculate the mutual information between Alice and Eve is
more involved. We expand 
\be
\ket{A} = \alpha_A \ket{00} +\beta_A \ket{10} + \gamma_A \ket{01} +\delta_A
        \ket{11} \ ,
\ee
where the complex coefficients have to satisfy
 \be
 \abs{\alpha_A}^2  +\abs{\beta_A}^2  + \abs{\gamma_A}^2  +
 \abs{\delta_A}^2 = 1
\ee       
and similarly for $\ket{B},\ket{C},\ket{D}$.
We are free to choose $\ket{B}$ as one of the four basis vectors, e.g.
$\ket{B}=\ket{00}$ and can fulfill the first constraint in 
equation (\ref{constraint1}) by setting $\ket{D}=\ket{11}$, without loss of
generality.   
We then find for the mutual information the form
\bea
I^{AE}= &1+ \half (&\tau[F\abs{\alpha_A}^2+(1-F),F\abs{\alpha_C}^2]+
         \nonumber \\
           & &    \tau[F\abs{\beta_A}^2,F\abs{\beta_C}^2]+\nonumber \\
           & &    \tau[F\abs{\gamma_A}^2,F\abs{\gamma_C}^2]+\nonumber \\
             & &  \tau[F\abs{\delta_A}^2+(1-F),F\abs{\delta_C}^2] ) \ ,
\label{muinf}
\eea
where we define
\be
\tau[x,y] = x\log x+y\log y -(x+y)\log(x+y) \ .
\ee
Note that $-\tau[x,1-x]$ is the entropy function. 
Equation (\ref{muinf}) is the mutual information which Eve reaches when
postponing the measurement until she learns which basis was used by
listening to the public channel. 
\par
The task is to maximize $I^{AE}$ with the constraints 
of eq. (\ref{constraint1}) and (\ref{constraint2}).
 The method of Lagrange multipliers leads to a set of
equations which can not be simultaneously
fulfilled unless $\alpha_A=\alpha_C=0$ and 
$\delta_A=\delta_C=0$. This means that the best solution for Eve is to use
states such that $ \bra{A} B\rangle =0 =\bra{C} D\rangle $, which one would
have expected. 
\par
\begin{figure}[hbt]
\setlength{\unitlength}{1pt}
\begin{picture}(500,300)
\epsfysize=10cm
\epsffile[72 230 540 560]{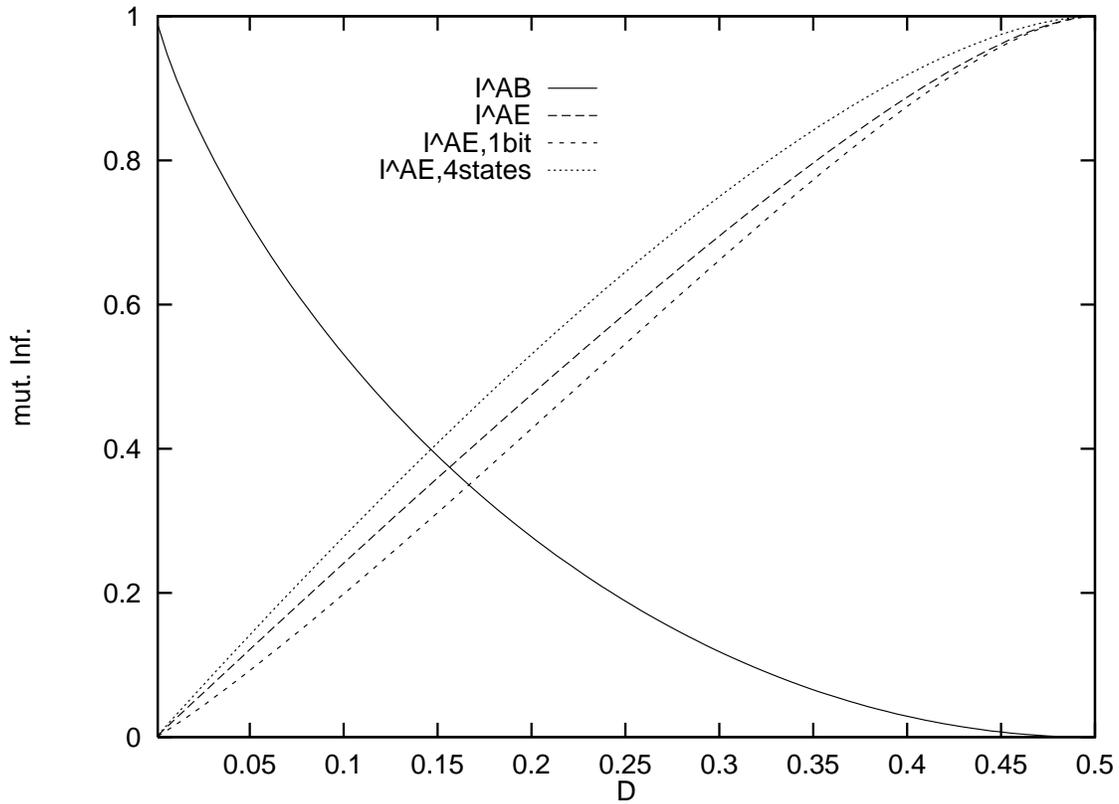}
\vspace{-0.2cm}
\end{picture}
\vspace{7cm}
\caption[]
        {\small Maximal mutual information $I^{AE}$ between Alice and Eve
               as function of Bob's disturbance $D$. The upper curve holds for
               BB84 \cite{fupe} and
               is shown for the purpose of comparison.
               The lower curves refer to the six-state
               protocol. Their analytic forms are shown in equations 
               (\ref{2info}) and (\ref{1info}).
               The mutual information between
               Alice and Bob is in both scenarios given by the curve $I^{AB}$.
               }
\label{figure1}
\end{figure}

 Now we have only two parameters left,  $\beta_A$ and $\beta_C$,
 and can
write
\be
I^{AE}= 1+ \half F(
             \tau[\abs{\beta_A}^2,\abs{\beta_C}^2]+
             \tau[(1-\abs{\beta_A}^2),(1-\abs{\beta_C}^2)] ) \ ,
\label{mu1inf}
\ee
which is a concave function. 
Here we have  used 
\be
\tau[Fx,Fy] = F\tau[x,y]\ .
\ee
It is  straightforward to write down the system of equations which has to
be fulfilled in order to maximize $I^{AE}$. Due to their high symmetry 
one   can find one solution easily, namely
\be
\abs{\beta_A}^2 = 1-\abs{\beta_C}^2\ ,
\ee
and thus
\be
I^{AE}  =  1+F\cdot \tau[\abs{\beta_A}^2,1-\abs{\beta_A}^2]\ .
\ee
By checking the higher derivatives one confirms that this is a maximum,
which is, due to concavity, the absolute maximum.
Inserting  into the second  line of eq. (\ref{constraint1}) allows us 
to find the `best' relative phase between $\ket{A}$ and $\ket{C}$
and thus leads
to the
solution
 for the highest mutual information that Eve can extract from measuring
her two qubits,
\bea
I^{AE}&=& 1+ (1-D) \left\{ f(D) \log f(D) + (1-f(D)) \log (1-f(D))
        \right\} \ , \nonumber \\
f(D) & = & \half \left( 1 + \frac{1}{1-D}\sqrt{D(2-3D)} \right) \  \ .
\label{2info}
\eea
This function is shown in figure \ref{figure1}, where we also give  
 the corresponding mutual
information for BB84, taken from~\cite{fupe}, 
for the purpose of comparison.
 ($I^{AB}$
is identical in both cases.) 
Note that our curve lies everywhere below
the one for the BB84 case.
The six state protocol is  therefore  
more secure against eavesdropping on
single qubits.
\par
In our case both bits of Eve carry mutual information, 
unlike the one described by
\cite{fupe}. If she would  either measure only one of her two bits,
or if she would  use a 1-bit probe from the beginning,
 her maximal 
information would be
\bea
I^{AE,1bit}&=& 1+ f_1(D) \log f_1(D) + (1-f_1(D)) \log (1-f_1(D)) \ , \nonumber \\
f_1(D) & = & \half \left( 1+D +\sqrt{D(2-3D)}\right) \ ,
\label{1info}
\eea
which is the lowest curve  in figure~\ref{figure1}. The calculation 
for the 1-bit
probe follows the same line as explained above for the 2-bit probe,
but is less involved.
 Note that
in order to maximize her mutual information in the six-state scheme Eve
necessarily needs two qubits as resource, whereas for BB84 a 1-bit probe is
sufficient to reach optimality \cite{niu}.
\par 
It is worth mentioning that the  optimal
unitary transformation which leads to equation (\ref{2info}) disturbs
 {\em all} Bloch vectors in the same way, not only
the six states used by Alice, and allows Eve to gain the same information in 
{\em all} possible bases. In other words: the optimal eavesdropping action 
for six states is shown to be a
 universal transformation.
 This means that using a bigger number of states
cannot increase security. The gain in security described in this paper is due
to the fact that the three bases  are spanning the full Bloch sphere,
as opposed to the case of BB84 where only a two-dimensional plane is spanned.
\par
The scheme described in \cite{bb84} 
 can also be realized by Alice and Bob sharing a singlet, 
i.e.  a maximally
entangled state. This was discussed in \cite{ekert,bennett}.
In this case, which we will consider for the rest of this article,
 Alice and Bob can test for eavesdropping by
calculating S, the correlation coefficient in the CHSH inequality.  
Without
any disturbance of Bob's bit
they will find $\abs{S}=2 \sqrt{2}$. 
This value is decreased when Eve
interacts unitarily with Bob's bit.
As  was shown in \cite{fupe}, the
intersection of the two curves for $I^{AB }$ and $I^{AE}$
 corresponds to $\abs{S}=2$, i.e. at
disturbances $D\geq \half(1-1/\sqrt{2})$
the CHSH inequality 
(between Alice  and Bob) is not violated. 
\par
The natural question arises whether the corresponding
 intersection for the generalized 
scheme is related to a generalized Bell inequality.  
In the six-state protocol the reduced density matrix of Alice and Bob after
Eve's interaction reads
\be
\rho^{AB} = 
\frac{1}{2}
\left( \begin{array}{cccc} D & 0 & 0 & 0 \\
                             0 & 1-D & 2D-1 & 0 \\
                             0 & 2D-1 & 1-D & 0 \\
                             0 & 0 & 0 & D \end{array}\right)
\ee
where the matrix elements are written in the order 00,10,01,11. For {\em any}
number of measurement directions  that Alice and Bob can use to test a
Bell inequality
we find
\be
\abs{S(D)}=\abs{S_q} \cdot (1-2D) \ 
\ee
where $S_q$ denotes the  correlation for $D=0$, i.e. the undisturbed
singlet. Thus  in our case
the measurement directions that are optimal for the singlet are
also optimal for $D\neq 0$, i.e. a mixed state. This does not hold in general
\cite{horo3}.
 We will refer to the disturbance where $\abs{S(D)}=\abs{S_c}$,
i.e. where $S$ reaches the classical limit,
as 
$D_c$.
\par
Let us first look at the case where Alice and Bob are using two measurement
directions each that do not necessarily lie in a plane. Here the 
 inequality for a model with local hidden variables
reads $\abs{S}\leq 2$.
 \par
We can make use of Cirel'son's inequality \cite{cirel} 
in which the norm of the operator
\be
C= \vec{a}_1 \cdot \vec{\sigma}^a \otimes \vec{b}_1 \cdot 
   \vec{\sigma}^b + 
 \vec{a}_2 \cdot \vec{\sigma}^a \otimes \vec{b}_1 \cdot 
   \vec{\sigma}^b +
    \vec{a}_2 \cdot \vec{\sigma}^a \otimes \vec{b}_2 \cdot 
   \vec{\sigma}^b - 
    \vec{a}_1 \cdot \vec{\sigma}^a \otimes \vec{b}_2 \cdot 
   \vec{\sigma}^b  \ \ 
\ee
is shown to obey $||C||\leq 2\sqrt{2}$.
(Here $\vec{a}_i$ refer to Alices directions of measurement and $\vec{b}_i$ to
those of Bob.)
This means that  the
maximal value the quantum correlation can take is $\abs{S_q}=2\sqrt{2}$, no matter
whether  the measurement directions span a plane or a sphere. This 
value is reached in the CHSH scenario. One can intuitively understand this in
the following way: in order to maximize the sum of scalar products of the
measurement directions  their relative angles have to be as small as possible,
i.e. they have to lie on a great circle of the sphere.
Thus we cannot find a  ratio for $\abs{S_q/S_c}$ that is higher than $\sqrt{2}$ 
and therefore  
we cannot establish a Bell
inequality in the sphere that corresponds to the intersection of  $I^{AE}$
with $I^{AB}$ for the generalized protocol, because here $D_c$ is larger than
in BB84. 
\par 
We can generally
exclude such a correspondence for  $n$  measurements
by each party,
i.e. chained Bell inequalities \cite{braun}: the inequality reads now
$\abs{S}\leq 2n-2$.
The relevant operator $C$ for this
case can be written as a sum of operators of the form used in Cirel'son's
inequality which we  call $C_1,...,C_{n-1}$. 
Due to the inequality
\bea
||C|| &= &||C_1 + C_2 + ... + C_{n-1}|| \leq
||C_1||+...+||C_{n-1}|| \nonumber \\
  & \leq &  (n-1)\cdot 2\sqrt{2}
  \eea
  we know an upper limit of the quantum correlation. Thus we find 
  $\abs{S_q/S_c}\leq \sqrt{2}$ as in the paragraph above and can generally exclude 
  the mentioned connection.
\par 
  Note that inequalities like the original Bell inequality and a 
  recent suggestion by
  Ardehali \cite{arde} where two directions of measurement coincide cannot be
  used for our purpose: the eavesdropping interaction causes the expectation
  value $\langle \vec{a} \cdot \vec{\sigma}^a \otimes \vec{a} \cdot 
   \vec{\sigma}^b \rangle$ to be smaller than 1 if $D>0$.
\par
To summarize, we have discussed a quantum cryptographic protocol based on six
quantum states and shown that it is safer against eavesdropping 
on single qubits than the BB84 scheme, 
because  Eve's maximal mutual information is
smaller than in the BB84 scenario. Furthermore, in order to reach the maximal
mutual information the eavesdropper needs to use a two-bit probe and thus has
to perform a more complicated transformation than in BB84. If her resource
consists of only one qubit, she gains even less information.
We have to mention some practical
disadvantage: in order to establish a key one
 will here loose 2/3 of the signals rather than 1/2 in the BB84 scenario, when
 using equal probabilities for all states.
 We have also shown that the best way to test a CHSH inequality is to use
 measurement directions that lie in a plane.
In the six-state protocol
 there is no natural relation between the classical  limit 
 of a Bell-type correlation coefficient and
 the intersection of the information curves.
We hope that 
this cryptographic
scheme may reach practical relevance in the light of recent suggestions to
produce maximally entangled pairs of distant atoms \cite{brie},
see also \cite{pavi}.
 \par  Valuable critics and  comments by Asher Peres
  and inspiring discussions with Alois W\"{u}rger  
 are gratefully acknowledged. 
 Questions by Charles Bennett and Chris Fuchs motivated
 the paragraph about the use of more than  six states.
  This work was supported  by 
the European TMR Research Network ERP-4061PL95-1412.

\newpage
\vspace{-2cm}

\end{document}